\documentclass[12pt]{iopart}

\usepackage{amssymb}
\usepackage{graphicx}
\usepackage{units}
\usepackage[separate-uncertainty=true]{siunitx}

\begin{document}

\title{Correlated photon-pair generation in a liquid-filled microcavity}

\author{Felix R{\"o}nchen, Thorsten F. Langerfeld,  and Michael K\"ohl}

\address{Physikalisches Institut, Universit\"at Bonn, Wegelerstrasse 8, 53115 Bonn, Germany}
\begin{indented}
\item[]22.7.2019
\end{indented}
\begin{abstract}
We report on the realization of a liquid-filled optical microcavity and demonstrate photon-pair generation by spontaneous four-wave mixing.
The bandwidth of the emitted photons is $\sim 300$\,MHz and we demonstrate tuning of the emission wavelength  between 770 and 800 nm.
Moreover, by employing a liquid as the nonlinear optical medium completely filling the microcavity, we observe more than a factor $10^3$ increase of the pair correlation rate per unit pump power and a factor of 1.7 improvement in the coincidence/accidental ratio as compared to our previous measurements.
\end{abstract}

\maketitle

\section{Introduction}

Nonclassical states of light, such as correlated photon pairs \cite{Hong1987,Bouwmeester1997} or anti-bunched photons, have contributed significantly to the tests of fundamental quantum mechanics and to interconnect remote quantum systems \cite{Kimble2008}.
One particular challenge when coupling different quantum systems is to match their wavelengths \cite{Siyushev2014, Meyer2015} or to bridge wavelength gaps.
Different methods from non-linear optics, such as spontaneous parametric downconversion (SPDC) and spontaneous four-wave mixing (SFWM), have been employed to generate correlated photon pairs.
SPDC sources with short crystals and SFWM sources, for example in optical fibers \cite{Fiorentino2002,Fan2005}, employ high-intensity, short-pulse pump lasers propagating through non-linear optical media.
The simplicity of these schemes results from the weak requirements regarding phase matching and leads to correlated photon pair emission in a broad bandwidth of several THz.
In contrast, narrow-band sources for photon pairs have often utilized non-linear media in optical cavities in order to enhance the field strength and control the emission bandwidth.
For a recent compilation of parameters see \cite{Savanier2016}.

Little attention has generally been devoted to the generation of photon pairs in liquid non-linear media \cite{Barbier2015}, and no experiments in optical cavities have been reported. In this work, we propose and demonstrate a novel approach of using a liquid-filled optical microcavity to prepare correlated photon pairs with tuneable wavelength separation by SFWM.
The four-wave mixing process absorbs two photons from the continuous-wave pump light field at frequency $\omega_0$ and produces photon pairs at frequencies $\omega_{n,\pm}=\omega_0\pm n\cdot \omega_{FSR}$, where $\omega_{FSR}=\frac{\pi c}{L}$ denotes the free spectral range of the Fabry-Perot cavity of length $L$, $c$ is the speed of light, and $n$ is an integer.
In principle, the usable range of $n$ is only limited by the bandwidth of the high-reflectivity coating of the cavity mirrors and the transparency of the liquid.
The liquid-filled approach of the optical cavity has significant advantages: (1) many liquids exhibit significantly higher non-linear refractive indices than solids, and (2) there is no additional interface between the non-linear medium and mirrors, which would affect the cavity performance.
The latter is an issue, in particular, for the highly-curved mirror of our microcavity.

The optical Kerr effect in solids has a very fast response time, typically, in the few femtosecond range or below.
This has been confirmed by experiments with attosecond laser pulses \cite{Sommer2016} and rapidly-oscillating optical fields \cite{Langerfeld2018}.
In liquids, however, the situation is more complex \cite{Boettcher1996,Righini1993}.
The optical Kerr effect has a contribution from both the purely electronic degree of freedom and the reorientation dynamics of the molecule if it has an anisotropic polarizability.
The latter has a complicated dynamics as it involves intermolecular interactions as well as vibrations and rotations.
There has been a long-standing tradition to study the optical Kerr effect in liquids using pump/probe schemes with adjustable time delay and it has been found that the peak Kerr response has delays with respect to the pump field in the few-ps range and that the instantaneous response is smaller by one or two orders of magnitude as compared to the peak response \cite{Zhao18}.

\begin{figure}[htbp]
	\centering
	\includegraphics[width=0.5\columnwidth]{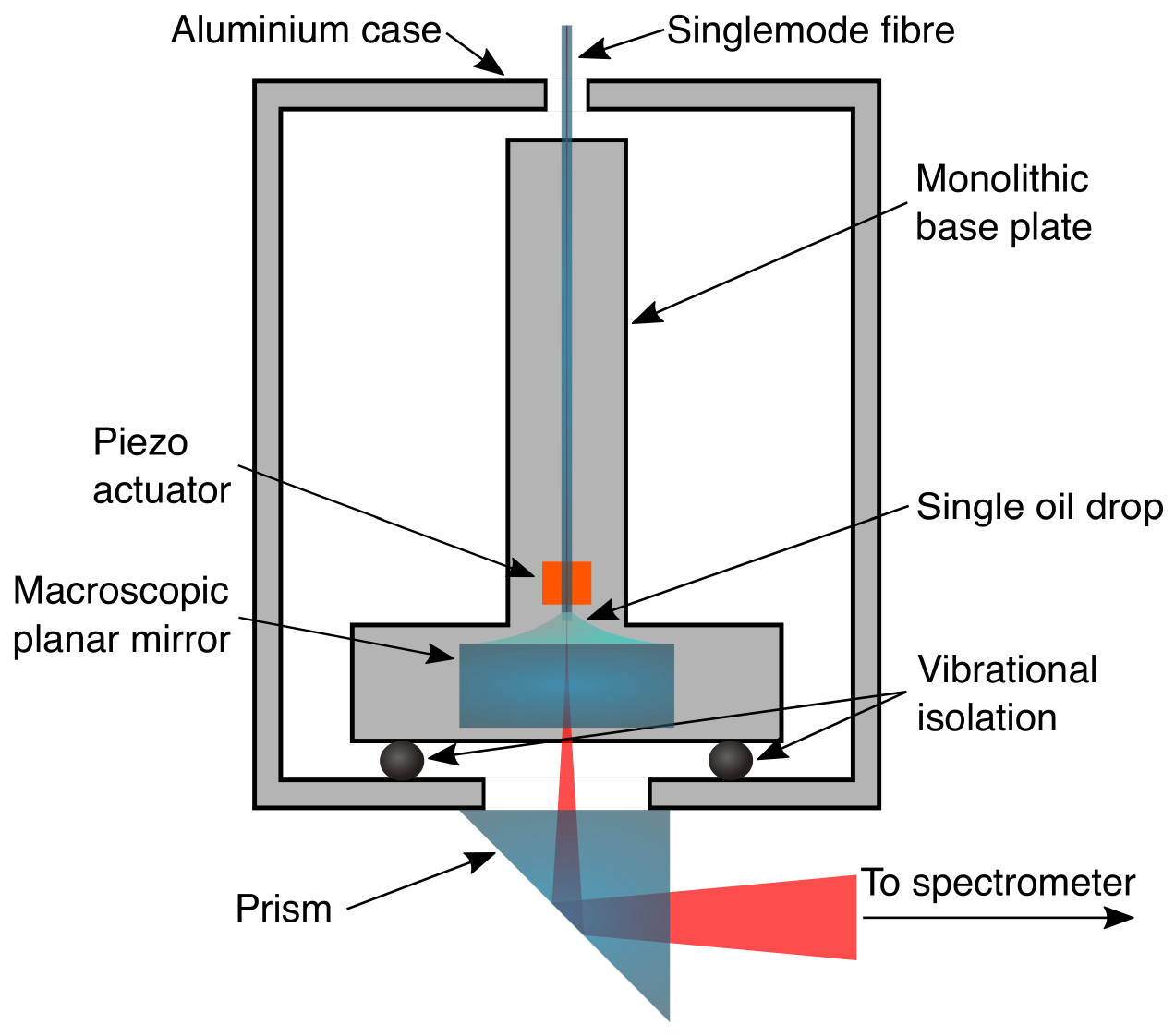}
	\caption{Schematic setup of the experiment.}
	\label{fig0}
\end{figure}

\section{Experiment}

We have constructed a Fabry-Perot microcavity composed of a micromachined and coated endfacet of an optical fiber as one mirror \cite{Hunger2010,Muller2010,Steiner2013,Brandstaetter2013,Takahashi2014,Uphoff2015,Gallego2016,Janitz2017,Langerfeld2018} and a conventional planar mirror with identical coating as the second mirror (see Figure 1).
The length of the cavity is $L=\SI{38.4}{\micro m}$ and the finesse is $F=\pi/(T+L)=12500\pm 500$ with a nominal mirror transmission of $T=100$\,ppm and intracavity losses of $L\simeq 100$\,ppm per mirror.
The radius of curvature of the fiber mirror is $R=\SI{200}{\micro m}$, giving rise to a $1/e^2$-beam radius on the planar mirror of $w_0=\SI{3.5}{\micro m}$.
This small mode waist enhances the desired nonlinear effects.

\begin{figure}[htbp]
	\centering
	\includegraphics[width=0.5\columnwidth]{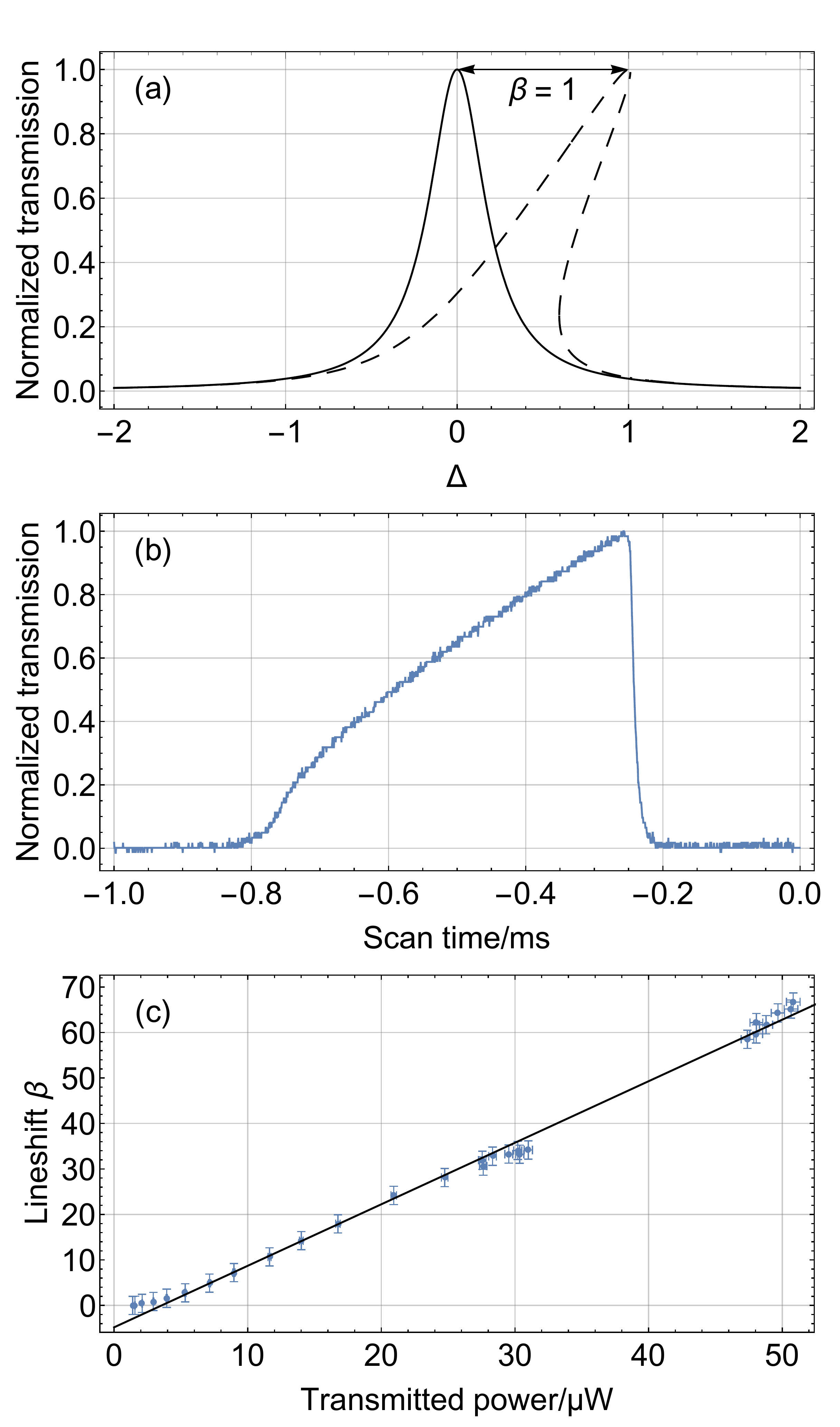}
	\caption{(a) Bistability of the lineshape function of a non-linear optical cavity [see equation (\ref{eqn1})]. (b) Measured bistable cavity lineshape. The lineshift $\beta$ is determined by the endpoints of the interval $\{x|A(x)>0.1A_\text{max}\}$, where $A$ is the amplitude with its maximum value $A_\text{max}$. To calibrate the time-axis into frequency units the resonance is measured and fitted for the lowest possible power, for which the lineshape is a symmetric Lorentzian of known width. (c) Measured lineshift $\beta$ for different powers. The linear dependence is expected from equation \ref{eqn1} and signals a constant refractive index inside the cavity.}
	\label{fig1}
\end{figure}

We have filled the cavity with the synthetic silicone oil (Tetramethyl-tetraphenyl-trisiloxane) with high optical transparency.
The refractive index of the oil leads to an increase of the optical path length of the cavity which has been detected by a change of the free spectral range from $ \SI{3.901\pm0.001}{THz} $ to $ \SI{2.507\pm0.002}{THz} $.
From this we have determined the refractive index of the oil as $n=1.556$ \cite{Gelest}.
However, the absorption coefficient of the oil has not been accurately determined previously.
After filling the microcavity with the oil, we have not observed a change of the cavity finesse.
Instead, the cavity linewidth  decreased from \SI{313}{MHz} to \SI{200}{MHz}.
Including the effect of the changing mirror reflectivity in presence of the oil, this sets an upper limit on the absorption coefficient of $\alpha=\SI{5}{m^{-1}}$, which is comparable to pure SiO$_2$ and indicates that the oil is a very high quality optical medium.

Strong pumping of the cavity with intracavity intensities of up to $10^{11}$\,W/m$^2$ leads to significant thermal effects.
For example, we observe the characteristic \cite{An1997,Dube1996,Langerfeld2018} bistable cavity line shape, which is a result of the optical path length change caused by absorption-induced heating from the high-intensity pump field inside the oil.
To lowest order, it can be modeled by an additional power-dependent detuning in the Lorentzian cavity lineshape \cite{An1997}.
\begin{equation}
	\frac{P_\text{out}}{P_\text{in}}=\frac{1}{(\Delta-\beta'P_\text{cav})^2+1}
	\label{eqn1}
\end{equation}
with the detuning $\Delta=\frac{\nu-\nu_\text{res}}{\delta \nu}$, the natural  cavity linewidth  $\delta \nu$, and the lineshift $\beta=\beta'P_\text{cav}$.
The lineshift broadens the resonance by several tens of the natural linewidth (see Fig. \ref{fig1}).
We measure the lineshift as a function of the transmitted power and observe a linear behaviour (see Fig. \ref{fig1}c) \cite{An1997}.
From this we deduce a constant outcoupling efficiency and finesse, even for increasing powers, and hence the intracavity power can be determined from the transmitted power.
Moreover, we estimate the temperature increase inside the cavity by connecting the resonance shift to the temperature increase $\delta T$ inside the cavity via
\begin{equation}
\beta = -\frac{\nu_\text{res}}{\delta\nu}\times\frac{\Delta L_\text{eff}}{L_\text{eff}}\approx  - Q\times \frac{C_\text{TO}}{n}\times \delta T
\end{equation}
with the quality-factor $Q=2\times10^6$, the thermo-optic coefficient $C_\text{TO}$ and the refractive index $n=1.56$.
The tabulated thermo-optic coefficients of various silicone oils are in the range of $C_\text{TO} \sim \SI{3e-4}{K^{-1}}$ \cite{VanRaalte1960}, however, the value for our specific oil is not known. Using the average value, we estimate a temperature increase of $\delta T=\SI{0.2}{K}$.
The comparatively low temperature increase results both from the low absorption coefficient and from the fact heat convection in liquids leads to a much faster heat dissipation than heat conduction alone, which would be the mechanism in solids.

In order to minimize variations resulting from thermal effects, we lock the cavity to the pump laser using a Pound-Drever-Hall locking scheme and as a result the residual  power fluctuations are below 2\%.
Using two additional lasers, we  measure the frequency difference to the higher frequency $(+)$ and lower frequency $(-)$ longitudinal modes for the same order $n$.
We then shift the center frequency of the cavity to the dispersion-compensated point such that both free spectral ranges are equal to within the cavity linewidth.
This dispersion compensation plays the role of fulfilling both the phase matching condition and the energy conservation.
Generally, we find that different pump power levels affect the dispersion compensation and we perform the compensation individually for each power.

The output of the Fabry-Perot cavity is spectrally dispersed by a home-built grating spectrometer, with a resolution of $\lambda/\delta\lambda=6000$ and the output is recorded with a pair of single photon counters (SPCM).
Additional dielectric line filters protect the SPCM from stray light.
Both line filters have a linewidth of 3\,nm and transmit light at the respective detection wavelengths while providing a suppression of the pump stray field on the SPCM of at least four orders of magnitude.
The measured quantum efficiencies of the two beam paths from the cavity are 9.9\% and 7.2\% including output coupling from the cavity and photon detection efficiencies.
We record the SPCM signals using a time-to-digital converter with timing resolution of 40\,ps, much better than the SPCM timing jitter of 350\,ps, and subsequently perform a correlation analysis with adjustable bandwidth.

\section{Results}

In Figure 2a we show a typical two-photon correlation measurement of order $n=2$, i.e., with photon frequencies of $\omega_{2,-}=2\pi \times 377.155$\,THz and $\omega_{2,+}=2\pi\times 387.155$\,THz for an intracavity  power of \SI{0.58}{W}.
The correlation signal shows a coincidence-to-accidental ratio (CAR) of $3.4\pm0.5$ and thus clearly signals the nonclassical nature of the emitted photon pairs \cite{Zou1991}.
The full width at half maximum (FWHM) of the correlation peak is $\SI{1.06\pm0.08}{ns}$, which corresponds to a cavity linewidth of $\SI{328\pm19}{MHz}$.
We attribute the increased width to a higher transmission of the dielectric coating at signal and idler frequencies than at the pump frequency.

In Figure 2b we show the scaling of the rate of photon pairs vs. intracavity power.
As expected for a spontaneous four-wave mixing process, we observe a quadratic dependence on pump power.
We compute the expected flux of photon pairs from SFWM from the optical cavity as \cite{Lamprecht1996}:

\begin{equation}\Gamma=\frac{\omega_{FSR}F}{\pi^2} \left[k\int n_2(z) I(z) \mathrm{d} z\right]^2.%=93000 \times n_2^2 \text{W}^{-2}\text{s}^{-1}.
\end{equation} 
The quantities $n_2(z)$ and $I(z)$ are the nonlinear refractive index and the light intensity, respectively, and $k=2\pi n/\lambda_\text{vac}$ is the wave vector of the light with the refractive index $n$.
The intensity is connected to the intracavity power via $I=\frac{P}{\pi w_0^2}$. 
Note that the quantity $n_2$ is proportional to the Kerr constant evaluated at an optical frequency  and hence will be lowered as compared to the tabulated dc ($\omega \to 0$) value of $n_2^{dc}=\SI{44e-20}{m^2W^{-1}}$ \cite{Ledzion1999} by the mechanisms discussed in the introduction. 
In principle, the integral over the Kerr coupling would even include the non-linear effects arising from the mirror coating \cite{Fedulova2016,Langerfeld2018}, however, they are approximately two orders of magnitude smaller and can be neglected. 
Comparing the theoretically expected flux $\Gamma$ to the experimentally determined rate coefficient $\Gamma_\text{exp}=\SI{1.12\pm0.05}{W^{-2}s^{-1}}$, see Figure 2b, we deduce the nonlinear refractive index  $n_2=\SI{3.62\pm0.31e-20}{m^2W^{-1}}$, which is one order of magnitude lower than the dc-value and comparable to SiO$_2$.

\begin{figure}[htbp]
	\centering
	\includegraphics[width=0.5\columnwidth]{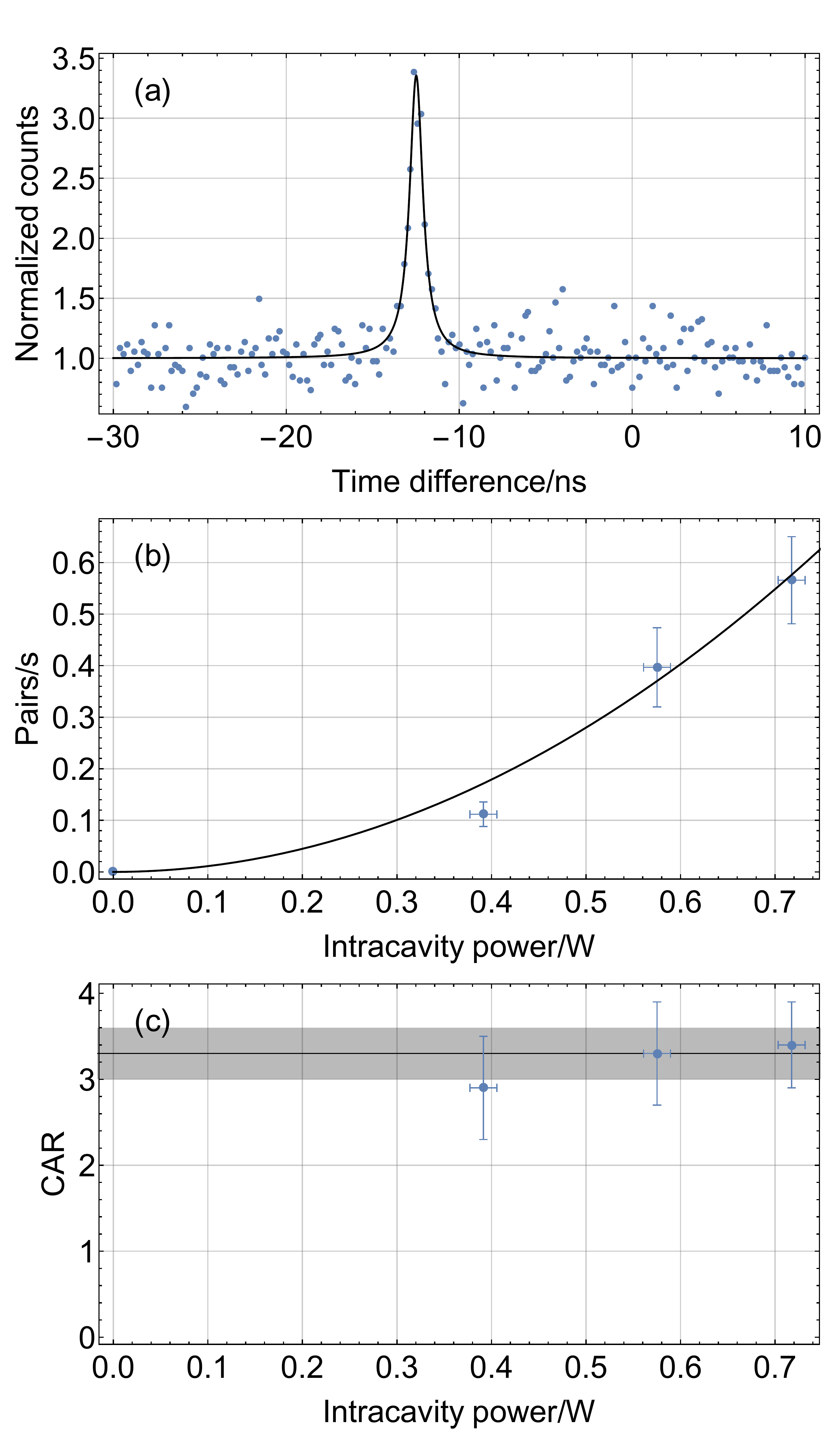}
	\caption{(a) Normalized coincidence signal of second order $(n=2)$ for a measurement time of 20\,min. The shifted peak-position at -12\,ns stems from a length difference of the detector cables. The solid line shows a Lorentzian fit to the data.	(b) Power dependence of the coincidence signal. The solid line is a quadratic fit yielding a curvature of $\Gamma_\text{exp}=1.12\pm0.05$\,W$^{-2}$s$^{-1}$.  (c) CAR-values corresponding to the data depicted in (b) together with the theoretically expected value, see text. }
	\label{fig2}
\end{figure}

On both spectrometer channels $(+n)$ and $(-n)$ we find a linear dependence of the count rate $R_{\pm n} = \gamma_{\pm n}\cdot P$ on the power coupled into the cavity.
We interpret this result as Raman scattering in the medium giving rise to background emission centered near the pump wavelength. The count rates are several orders of magnitude above the detector dark count rates and they limit the CAR to $\frac{2}{\pi \tau_c}\left(\frac{\Gamma_{\text{exp}}}{\gamma_1 \gamma_2}\right)$.
We have extracted the formula by assuming a Cauchy distribution of the correlated photon pairs of width $\tau_c$ and a binning time much shorter than the correlation time, which is well fulfilled in our experiment.
This shows that better background subtraction by spectral filtering at higher orders $n$ facilitates detection of correlations with higher CAR.
For $n=2$ we compute a CAR of $3.3\pm0.3$, which is in excellent agreement with the measured values depicted in \ref{fig2}c.

We also have observed correlated photon pairs in the third order at frequencies $\omega_{3,-}=2\pi \times \SI{374.831}{THz}$ and $\omega_{3,+}=2\pi\times \SI{389.988}{THz}$ for $\omega_0=2\pi \times 382.410$\,THz and \SI{0.67}{W} intracavity power.
This signals the versatility of our approach to generate photon pairs at controllable frequency difference.
In this context, the liquid filling of our cavity offers the unique advantage that continuous changes of the cavity length and hence continuous adjustments of the frequency of the emitted photon pairs are possible.
The measured photon rate coefficient in third order is \SI{0.58\pm0.22}{W^{-2}s^{-1}}.
In principle, the rate of photon pairs should be independent of the order of the free spectral range.
However, in our realization, we have observed some optical damage from the high-intensity experiments in the second order in addition to the lower reflectivity for increasing order of $n$, with both effects reducing the finesse.

In conclusion, we have demonstrated a  liquid-filled microcavity at very high finesse and show that liquids, in addition to their high nonlinearities, can exhibit very low absorption coefficients (comparable to pure SiO$_2$),  which makes them attractive for non-linear optics studies. We have studied the generation of correlated photon pairs from such a microcavity using spontaneous four-wave mixing. Photons are emitted in pairs of longitudinal modes equally split from the pump laser frequency and we have detected correlated pairs up to order $n=3$ thereby demonstrating a photon pair source with adjustable frequency spacing between the photons of a pair. 

This work was supported by DFG (SFB/TR 185, A2), Cluster of Excellence Matter and Light for Quantum Computing (ML4Q) EXC 2004/1 – 390534769, BMBF (FaResQ), and the Alexander-von-Humboldt Stiftung.

\end{document}